\newcommand{\ket}[1]{|#1\rangle}

\documentclass[prl,twocolumn]{revtex4}
\usepackage{graphicx}
\begin{document}
\title{Interference enhanced polarization entanglement and the concept of an Entangled-Photon Laser}
\author{Antia Lamas-Linares, Christian Mikkelsen, John C. Howell and Dik Bouwmeester}
\affiliation{
Centre for Quantum Computation, Clarendon Laboratory, University
of Oxford,\\ Parks Road, OX1 3PU Oxford, United Kingdom}
\date{\today}
\begin{abstract}
Inspired by laser operation, we address the question of whether
stimulated emission into polarization entangled modes can be
achieved. We describe a state produced by stimulated emission of
the singlet Bell state and propose a setup for creating it. As a
first important step towards an entangled-photon laser we
demonstrate experimentally interference-enhanced polarization
entanglement.
\end{abstract}
\maketitle
In the past two decades there has been a wealth of fascinating
experiments on the production and applications of entangled
photons. Whereas the earlier research mainly focused on fundamental
tests of quantum mechanics
\cite{Clauser,Aspect2,OuMandel,Shih,Kwiat1}, the recently emerged
field of quantum-information science added an unexpected twist to
the research leading to intriguing topics such as
entanglement-based quantum cryptography
\cite{Ekert1,Ekert2,Jennewein,Tittel}, and quantum teleportation
\cite{Bennett1,Popescu,Bouwmeester1,Boschi}. See
Ref.~\cite{Springer} for an overview and
Refs.~\cite{Brendel,Weihs,Pan1} for recent fundamental tests of
quantum mechanics. So far, mainly pairs of entangled photons,
produced by parametric down-conversion inside non-linear crystals
\cite{OuMandel,Shih,kim:00,Kwiat1,Kwiat2,Kwiat3,Atature} have been
used in the experiments and only few experiments addressed three-
and four-photon quantum correlations
\cite{Pan1,Pan2,Bouwmeester2,Ou,Martini1}. At the other end of the
optical energy range, where the number of photons is large and the
optical fields can be described by in- and out-of-phase amplitudes
with respect to a classical electro-magnetic field, quantum
correlations may still play an important role. For example,
squeezed light fields \cite{Slusher,Wu} can be used to create
entangled light beams (entangled in the noise-level of the
quadrature components ~\cite{Ou2} and to implement quantum
teleportation \cite{Vaidman,Braunstein,Furusawa}.

The nature of the quantum correlations at the few photon level and
at the high-intensity level are very different. For entangled
photons the quantum correlation is between degrees of freedom for
each particle (e.g. their polarization degrees of freedom). On the
other hand, the quantum correlation at the high-intensity level is
between photon numbers in the entangled modes.

In this Letter, we introduce the notion of laser-like operation for
polarization entangled photons. The quantum correlations will be
{\em both} between the polarization degree of freedom for each
particle and between the photon numbers in the entangled modes.
Therefore, the proposed device will operate in the new and
virtually unexplored regime of many-particle entanglement.

As the acronym LASER (Light Amplification by Stimulated Emission
of Radiation) indicates, entangled laser operation would mean that
a (spontaneously created) photon pair in two
polarization-entangled modes stimulate, inside a non-linear gain
medium, the emission of additional pairs. For this process a
non-linear gain medium is required for which we consider type-II
parametric down conversion \cite{Kwiat2}. A simplified interaction
Hamiltonian \cite{Kok,Simon} for the nonlinear interaction between
a classical pump field and two polarization-entangled modes is
given by
\begin{equation}
\hat{H}_{\mbox{\small{int}}}=e^{i\phi} \kappa  \hat{K}^\dagger
+e^{-i\phi} \kappa \hat{K} \,, \label{Hamiltonian}
\end{equation}
where \( \hat{K}^\dagger\equiv
(\hat{a}^\dagger_h\hat{b}^\dagger_v-\hat{a}^\dagger_v\hat{b}^\dagger_h)
\) and \(\hat{K}\equiv (\hat{a}_h\hat{b}_v-\hat{a}_v\hat{b}_h)\,,
\) are the creation and annihilation operators of polarization
entangled photon pairs in  modes $a$ and $b$. Horizontal and
vertical polarization are represented by $h$ and $v$, and $\kappa$
is a real-valued coupling coefficient. When acting on the vacuum
state the time evolution operator \(\hat{U}=\exp
({i\hat{H}t/\hbar}) \) yields  \cite{Kok}
\begin{equation}
\hat{U}(\tau)|0\rangle=e^{-q}\sum_{n=0}^{\infty}r^n \left(\sum_{m=0}^n(-1)^m|n-m,m;m,n-m\rangle \right) \,,
\label{eq:PDCstates}
\end{equation}
where $\tau\equiv \frac{\kappa t}{\hbar}$, $r\equiv\tanh\tau $ and
$ q\equiv 2\ln(\cosh\tau)$. We used the shorthand notation
$|i,j;k,l\rangle$ for $|i\rangle_{ah} |j\rangle_{av} |k\rangle_{bh}
|l\rangle_{bv}$ and $|0\rangle$ represents $|0,0;0,0\rangle$. The
fact that for a given $n$, all terms in Eq.~(\ref{eq:PDCstates})
have equal absolute amplitudes is the result of stimulated
emission. This is in contrast to the state obtained by the product
of $n$ entangled pairs in the $\Psi^-$ state, which would yield a
binomial distribution over the terms. An important consequence of
the equally weighted terms for a given $n$ is that they form a
highly polarisation entangled state of $2n$ photons. In addition
the state (\ref{eq:PDCstates}) is entangled in photon numbers in
the two modes, though only weakly since the terms with different
$n$'s have different weights $r^n$ ($r<1$).

The state described in Eq.~(\ref{eq:PDCstates}) would be a
desirable output for what we call a polarization entangled-photon
laser. It has similar features to the output of a conventional
laser, in the sense that the photon number distribution broadens
and shifts its peak as the average number of photons increases in
marked contrast to the behaviour of a thermal or squeezing
distribution. It does not however approach a stationary state,
because no loss mechanisms are included in the description. In a
conventional laser the intra-cavity photons can be seen as copies
of one another (except for the one spontaneously emitted photon in
the mode), which implies that a conventional laser can operate with
continuous outcoupling. In contrast, the photons in a polarization
entangled-photon laser form one complex entangled state. Therefore
the entangled fields first have to build up inside a high finesse
resonator and then they have to be switched out. This requires a
pulsed laser operation.

At the heart of our design for an entangled photon laser is the
bow-tie cavity, shown in Fig.~\ref{pro}(a). The reason for this
design is two-fold. First, since we consider non-collinear
down-conversion the modes diverge out of the gain region and have
to be redirected into the non-linear interaction region in order to
achieve efficient stimulated emission. Second, the two entangled
modes are counter propagating modes in the same resonator and
therefore experience exactly the same path length and mode
structure. By including a $\lambda/2$ wave plate in the resonator,
such that the vertical and horizontal field components will
interchange every round trip, each photon will also experience
exactly the same birefringence and spatial walk off as any other
photon. Therefore, the bow-tie ring resonator together with the
$\lambda/2$ wave plate guarantees the indistinguishability of all
photons created throughout the amplification process which is the
key to obtaining optimal entanglement. The resonator has to be
interferometrically stable with respect to the phase of the pump
laser pulses because the parametric down-conversion can be
considered as a phase-dependent gain mechanism.

There are many experimental complications to be overcome before an
entangled photon laser can be built, but as a first experimental
feasibility study of the main operation mechanism we reduced the
design to the double-pass configuration shown in Fig.~\ref{pro}(b).
Photon pairs in the singlet $\Psi^-$ Bell-state can be created in
the first and second pass of a noncollinear type-II parametric
down-conversion crystal. Since the creation is rather inefficient,
we restrict our attention to the observation of resonant
enhancement of the spontaneous emission rate which is an adequate
measure of how well the indistinguishability requirements can be
met.

The time-evolution operator for the two-pass system is given by
\begin{equation}
\hat{U}=\hat{U}_2 \hat{U}_1 =
e^{\tau(e^{i\theta}\hat{K}^\dagger_{2}-e^{-i\theta}\hat{K}_{2})}e^{\tau(\hat{K}^\dagger_{1}-\hat{K}_{1})}
\end{equation}
where $\theta=\phi + \pi/2$ and the subscripts on $\hat{K}$ denote
on which pass they operate. Expanding to first order yields
\begin{equation}
\hat{U}=1+\tau \hat{K}^\dagger_{1} +
e^{i\theta}\tau\hat{K}^\dagger_{2}
\end{equation}
where the terms containing $\hat{K}$ are neglected, since they
yield zero when they operate on the vacuum.  If the entangled
photon pairs created by $\hat{K}^\dagger_{1}$ and
$\hat{K}^\dagger_{2}$ are indistinguishable, then
\begin{equation}
\hat{U}=1+\tau \hat{K}^\dagger (1 + e^{i\theta}) \,. \label{theta}
\end{equation}
Equation (\ref{theta}) is most easily understood as an
interference pattern of two polarization-entangled photon-pair
amplitudes. Generalizing for $n$ passes, the result is that the
probability for obtaining a pair of entangled photons scales as
$n^2$ for indistinguishable fields and as $n$ for distinguishable
fields.

In addition to the general design of the counter-propagating
entangled-mode configuration and the $\lambda/2$ wave-plate, two
1mm BBO crystals, rotated by 90$^{\circ}$ with respect to the main
BBO crystal, and narrow-bandwidth filters f1 and f2 have been
included in the detection stage in order to complete the
birefringence and temporal walk-off compensation
\cite{Kwiat2,Zukowski}.

Figure \ref{result1}(a) shows the detected number of
polarization-entangled photon pairs as function of the difference
in path length between the entangled-photon feed-back loop and the
delay of the reflected pump-pulse. Figure~\ref{result1}(b) shows
that inside the region of interference, the coincidence rate
oscillates with a period corresponding to half the wavelength of
the pump field~\cite{milonni:96}.

The entanglement was assesed by looking at the visibility of the
coincidence rate variation as function of the difference in angle
between the polarizers in front of detectors D1 and D2 (see Fig.
\ref{vis}). In particular we fix one of the polarizers at $45^0$
and thus avoid the experimentally favoured basis $H$, $V$. This
visibility was measured both in the maximum overlap and in the no
overlap region with results of 0.92 and 0.83 respectively. The
higher quality obtained when the two passes are indistinguishable
can be explained as a by-product of the exchange in polarization in
the feedback loop. This has the additional effect of eliminating
the spectral distinguishability \cite{branning:99} between the
down-converted photons. In conclusion, the high interference
contrast indicates that the experimental setup operates under the
important indistinguishability requirements between pairs created
at different times.

Related experiments on interference enhanced emission and
stimulated emission of photon pairs that are not entangled in
polarization have been reported and analysed in
Refs.~\cite{Ou,Herzog,milonni:96}. Experiments and theory involving
single-photon injection into type-II crystals (rather than
pair-photon injection as presented in this letter) have been
reported in Ref.~\cite{Martini1,Martini2}.
%\cite{Simon}

Having demonstrated the feasibility of one of the main mechanisms
of an entangled photon laser we now briefly discuss what the
prospects are for the construction and possible applications of an
operational device. First of all, just like squeezing experiments,
the device will be very sensitive to loss, both in the generation
of the entangled output modes as well as during the propagation and
detection of the output state. With current technology it is
possible to obtain a round-trip cavity loss of 2-3\% (including
switching elements) and a pair-photon creation efficiency of 0.5
per pump pulse per pass of the nonlinear material. Under these
conditions polarization-entangled output modes with 10 to 1000
photons per pulse are conceivable. With the current advances in
ultra low-loss optical fibres and high efficiency multiphoton
detectors ~\cite{kim:99} the transmission and analysis of the
produced state might not be too daunting. However, despite
ingeneuos experimental progress, we will always have to face the
situation of losing photons in the process of creating,
transporting and analyzing the desired state (2). Since the desired
state is one large complex entangled state (entangled in
polarization and photon numbers) one might think that loss destroys
all the intersting properties of the state. However, on the
contrary, the complex entanglement has the remarkable feature that
the loss or measurement of one (or more) particles does not
eliminate all the entanglement between the remaining particles. To
illustrate this important property let us focus on the
(un-normalised) four-particle terms in Eq (2):
$\ket{2,0;0,2}-\ket{1,1;1,1}+\ket{0,2;2,0}$. Consider the case that
a polarisation measurement in the H, V basis is made on one
particle, say in mode $a$, and the measurement result is $h$. The
state of the remaining three  particles will be
$\sqrt{2}\ket{1,0;0,2}
-\ket{0,1;1,1}$ which contains (non-maximally) three-particle
entanglement. This feature of robustness seems to indicate a
connection with
\emph{high entanglement persistency} states \cite{Briegel}.

In this letter we introduced the concept of an entangled photon
laser and studied crucial experimental aspects of its construction.
Currently we are exploring the multi-particle correlations of the
entangled photon laser for novel secure quantum communication,
quantum cloning ~\cite{Simon}, and entanglement purification
~\cite{Geze}.

\noindent {\bf Acknowledgements}

\noindent We thank Piero Varisco and William Irvine for their experimental support. This
work was supported by the EPSRC GR/M88976 and the European QuComm
(ISI-1999-10033) projects.

\begin{figure}
\caption{The basic design of a polarization entangled-photon laser
is shown in (a). A linear pulsed pump resonator overlaps inside a
nonlinear crystal with two counter-propagating polarization
entangled modes. After the intra-cavity fields have built up during
several roundtrips the total field is coupled out by two optical
switches. Reducing the design to a double pass of the crystal leads
to setup (b). A frequency-doubled mode-locked Ti:Sapp laser (80 MHz
rep. rate, $\lambda_0=780$ nm) pumps a 2 mm BBO crystal. Pinholes
$p$ perform spatial selection of the entangled modes. The pump is
reflected onto itself by mirror M3 which is mounted on a
computer-controlled translation stage. Photon pairs are observed by
coincidence-detection between avalanche photodiodes D1 and D2 after
passing through polarizers, Pol1 and Pol2, and 5nm bandwidth
filters f1 and f2.} \label{pro}
\end{figure}
\begin{figure}
\caption{Graph (a) shows the number of detected photon pairs around the region of overlap between the
reflected pump and the photon-pair wavepacket from the first pass.
Graph (b) is a fine scan of the zero delay region. An average over
many oscillations gives us a period of $200\pm10$nm which
corresponds with the expected value of half the wavelength of the
pump field $\lambda_p/2=195$nm.}
\label{result1}
\end{figure}
\begin{figure}
\caption{(a), (b), (c) and (d) show the interference fringes for different relative orientations of the polarizers in front of the two detectors. In particular, one polarizer was fixed at $45^0$ while the other was varied from $45^0$ (a), through $75^0$ (b) and $105^0$ (c) to $135^0$ (d). Graph (e) shows the average of the maxima of the interference as a function of the relative angle. The visibility of this curve is $0.92\pm 0.02$) which is a direct measurement of the entanglement in the overlap region.} \label{vis}
\end{figure}

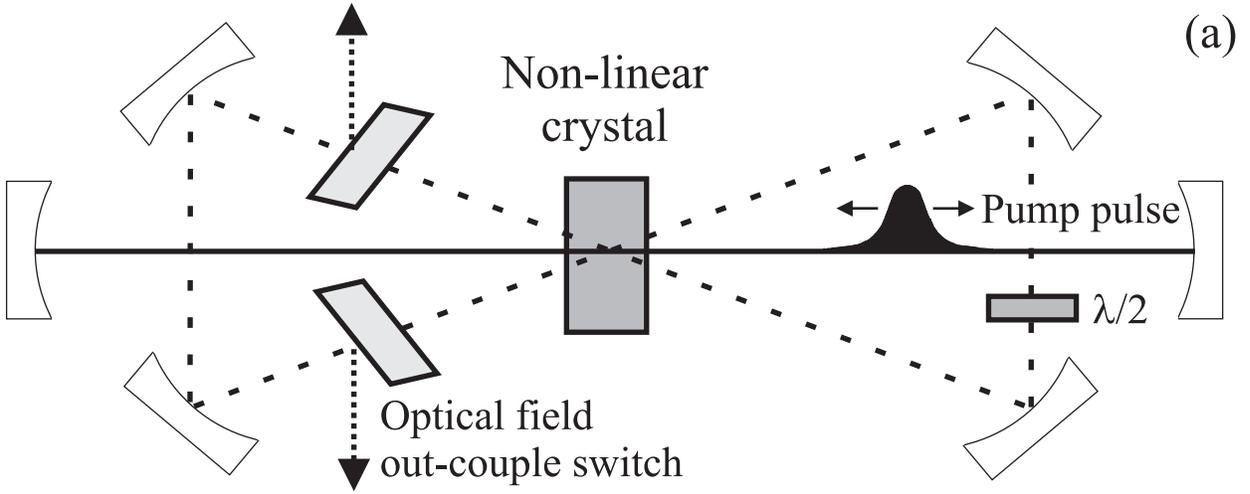
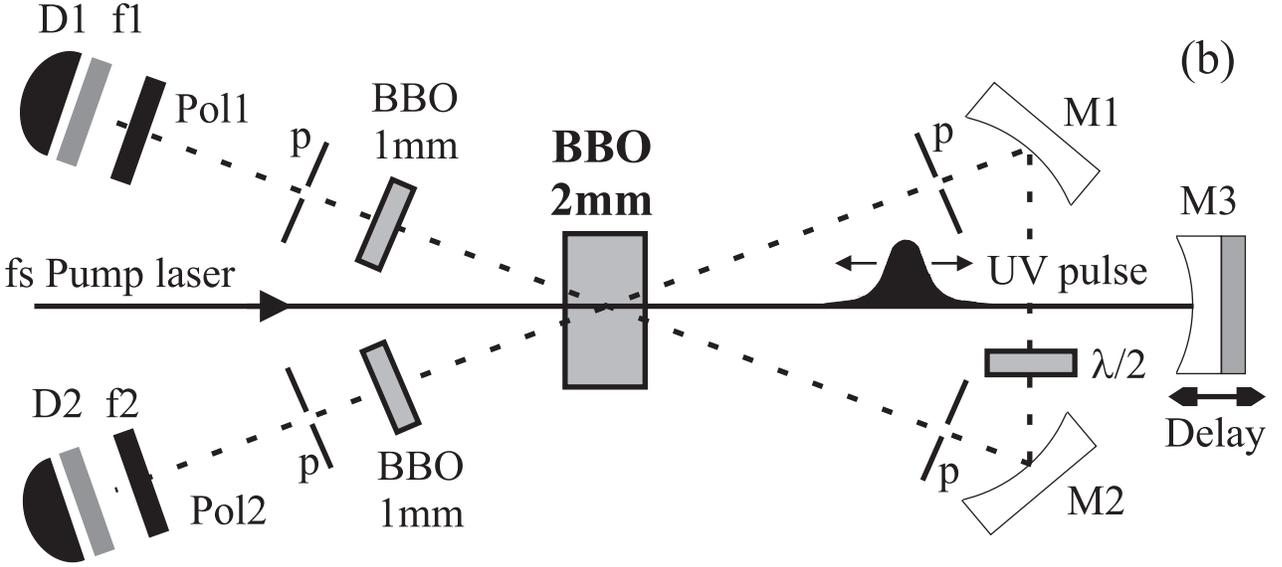
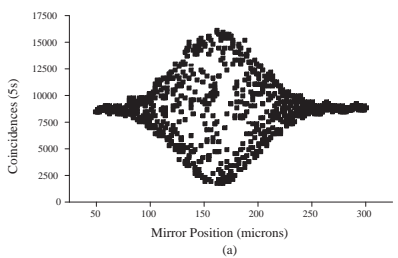
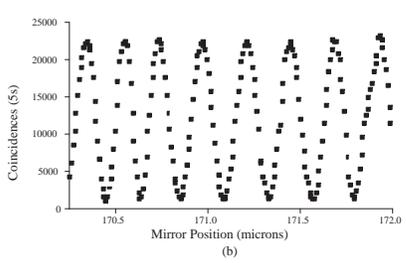
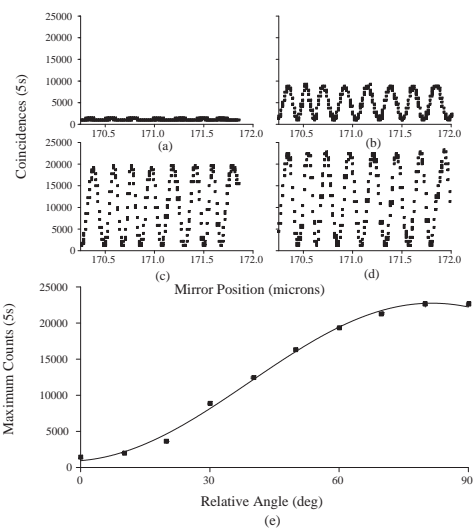
\end{document}